\documentclass[11pt]{article}

\columnwidth\textwidth
\usepackage{graphicx}
\usepackage{epsfig}

\begin{document}
\title{Coulomb distortion of relativistic electrons in the\\
nuclear electrostatic field}
\author{Andreas Aste, Cyrill von Arx, Dirk Trautmann\\
Department for Physics and Astronomy, University of Basel,\\
4056 Basel, Switzerland}
\date{October 3, 2005}

\maketitle

\begin{center}
\begin{abstract}
Continuum states of the Dirac equation are calculated numerically for
the electrostatic field generated by the charge distribution
of an atomic nucleus. The behavior of the wave functions of
an incoming electron with a given asymptotic momentum in
the nuclear region is discussed in detail and the results are
compared to different approximations used in
the data analysis for quasielastic electron scattering
off medium and highly charged nuclei.
It is found that most of the approximations
provide an accurate description of the electron wave functions
in the range of electron energies above 100 MeV typically used
in experiments for quasielastic electron scattering off nuclei
only near the center of the nucleus. It is therefore necessary
that the properties of exact wave functions are investigated in
detail in order to obtain reliable results in the data analysis of
quasielastic  $(e,e'p)$ knockout reactions
or inclusive quasielastic $(e,e')$ scattering. Detailed
arguments are given that the effective momentum approximation
with a fitted potential parameter is a viable method for
a simplified treatment of Coulomb corrections for certain
kinematical regions used in experiments.
Numerical calculations performed within the framework of the single
particle shell model for nucleons lead to the conclusion that our
results are incompatible with calculations performed
about a decade ago, where exact electron wave functions were used
in order to calculate Coulomb corrections in distorted wave Born
approximation.
A discussion of the exact solutions of the Dirac equation
for free electrons in a Coulomb field generated by a point-like
charge and some details relevant for the numerical
calculations are given in the appendix.
\end{abstract}
\end{center}

\noindent 

\section{Introduction}
Quasielastic $(e,e'p)$ knockout reactions
provide a powerful possibility to obtain information on
the electromagnetic properties of nucleons embedded in the
nuclear medium, since the transparency of the nucleus with respect to
the electromagnetic probe makes it possible to explore
the entire nuclear volume.
Inclusive $(e,e')$ scattering, where only the scattered electron is observed,
provides information on a number of
interesting nuclear properties like, e.g., the nuclear Fermi momentum
\cite{Whitney74}, high-momentum components in nuclear wave
functions \cite{Benhar94a}, modifications of nucleon form factors
in the nuclear medium \cite{Jourdan96a}, the scaling properties
of the quasielastic response allow to study the reaction
mechanism \cite{Day90}, and extrapolation of the quasielastic
response to infinite nucleon number
$A=\infty$ provides us with a very valuable observable
of infinite nuclear matter \cite{Day89}.
There is now considerable theoretical and experimental interest 
in extracting longitudinal and transverse structure functions as a function 
of energy loss for fixed three momentum transfer for a range of 
nuclei \cite{jlab}.
In August 2005, the Thomas Jefferson National Accelerator Facility (TJNAF) Proposal
E01-016, entitled `Precision measurement of longitudinal and transverse response
functions of quasi-elastic electron scattering in the momentum transfer range
$0.55$ GeV $\le$ $|\vec{q}|$ $\le$ $1.0$ GeV' was approved
such that the experiments will be performed in the near future at
the TJNAF using $^4$He, $^{12}$C, $^{56}$Fe and $^{208}$Pb as target nuclei.

The plane wave Born approximation
is no longer adequate for the calculation of scattering
cross sections in the strong and long-range electrostatic field of
highly charged nuclei, and it has become clear
in recent years that the correct treatment
of the Coulomb distortion of the electron wave function due
to the electrostatic field of the nucleus is unavoidable
if one aims at a consistent interpretation of experimental data.
E.g., it is still unclear whether the Coulomb sum rule is violated
in nuclei \cite{Morgenstern}.

Distorted wave Born approximation (DWBA) calculations with exact Dirac wave functions
have been performed by Kim {\emph{et al.}} \cite{Jin1} in the Ohio group
and Udias {\emph{et al.}} \cite{Udias,Udias2} for quasielastic
scattering off heavy nuclei. However, these calculations are cumbersome
and difficult to control by people who were not directly involved
in the development of the respective programs. Early DWBA calculations for
$^{12}$C and $^{40}$Ca were presented in \cite{CoHei}

Various approximate treatments have been proposed in the past
for the treatment of Coulomb distortions
\cite{Lenz,Knoll,Giusti,Rosenfelder,Giusti88,Traini88,Traini95,
Rosenfelder80},
and there is an extensive literature on the so-called eikonal approximation
\cite{Sucher,Blankenbecler,Wallace1,Wallace2,Abarbanel,Aste1,Aste2}.
At lowest order, an expansion of the electron wave function in
$\alpha Z$, where $\alpha$ is the fine-structure constant and $Z$
the charge number of the nucleus, leads to the well known
effective momentum approximation (EMA) \cite{Yennie64}, 
which plays an important role in experimental data analysis
and which will be explained below.

The effect of the charged nucleus on the electron wave function
is twofold:
Firstly, the (initial and final state) electron momentum $\vec{k}_{i,f}$
is enhanced in the vicinity of the nucleus due to the attractive
electrostatic potential, i.e., the wave length of the electron
is becoming shorter near the nucleus.
Secondly, the attractive potential of the nucleus leads to a
focusing of the electron wave function in the nuclear region.
Solutions for the Dirac equation for the scattering of electrons
in the nuclear field can be obtained from a partial wave
expansion, where the radial Dirac equation must be solved
numerically for each partial wave. To avoid such a computational
effort, an approximate treatment is often adopted, based on
a high-energy expansion in inverse powers
of the electron energy \cite{Lenz,Knoll,Rosenfelder}.
The resulting expression for the distorted electron wave function
is then expanded in powers of $\alpha Z$ as (we use units with
$\hbar=c=1$ throughout)
\begin{equation}
\psi_\tau=e^{\pm i \delta_{1/2}} \frac{k'}{k} e^{i \vec{k}' \vec{r}}
[1+g^{(1)} (a,b,\vec{k}',\vec{r}) + g^{(2)} (a,b,\vec{k}',\vec{r})+...]
u_\tau, \label{expansionalpha}
\end{equation}
where the sign $\pm$ refers to the two scattering solutions with
outgoing or incoming spherical waves, respectively, whereas
the corresponding indices $i,f$ are neglected
for the sake of notational simplicity,
$u_\tau$ is the plane wave spinor for
the electron with a given helicity $\tau$ and $\vec{k}$ is
the asymptotic electron momentum.
For the case of a uniform spherical charge distribution of radius $R$,
the values of $\vec{k}',a,b$ and $\delta_{1/2}$ are given by
\cite{Giusti88}
\begin{displaymath}
k'=k+ \frac{3\alpha Z}{2R} , \quad
\delta_{1/2}=\alpha Z \biggl(\frac{4}{3}-\log 2 k R \biggr)+b,
\end{displaymath}
\begin{equation}
a=-\frac{\alpha Z}{6 k' R^3}, \quad b=-\frac{3 \alpha Z}{4 k'^2 R^2},
\label{parameters}
\end{equation}
with $\vec{k}'$ parallel to $\vec{k}$. These values enter the
first-order term according to
\begin{equation}
g^{(1)}=a r^2 +iar^2 \vec{k}' \vec{r} \pm
ib [(\vec{k}' \times \vec{r})^2 + 2i \vec{k}' \vec{r}-
\vec{s} (\vec{k}' \times \vec{r})], \label{firstorder}
\end{equation}
where the spin operator $\vec{s}$ describes spin-dependent
effects which are comparably small for higher electron energies.
The meaning of the parameter $a$ can be easily understood
from a semiclassical observation.
For a highly relativistic electron with mass $m$
falling along the $z$-axis
(i.e. with zero impact parameter and $k \gg V \gg m$)
towards the nuclear center, the momentum inside the
spherical charge distribution is given by
\begin{equation}
{\tilde{k}}(z)=k-V(z), \quad 
V(r)=-\frac{3 \alpha Z}{2 R} +\frac{\alpha Z}{2 R} \biggl(
\frac{r}{R} \biggr)^2 \label{potmom}
\end{equation}
where $V(r)$ is the electrostatic potential inside the charged sphere.
Modifying the electron plane wave phase $e^{i \vec{k} \vec{r}}=e^{ikz}$
to $e^{i k' z (1 + a z^2)}$, as it is induced at lowest order by
the second term in $g^{(1)}$, leads to the $z-$dependent
electron momentum ${\tilde{k}}(z)$
\begin{equation}
\frac{1}{i} \frac{d}{dz} e^{i k' z (1 + a z^2)}=
k'(1+3az^2) e^{i k' z (1 + a z^2)}={\tilde{k}}(z)  e^{i k' z (1 + a z^2)}
\label{eiko1}
\end{equation}
in agreement with eq. (\ref{potmom}).
The parameter $b$ describes mainly the deformation of the wave front
and can also be derived from semiclassical observations.

The standard method (in the case of light nuclei) to handle Coulomb
distortions for elastic scattering in data
analysis is the effective momentum approximation (EMA), which corresponds
to the lowest order description of the Coulomb distortion in
$\alpha Z$.
EMA accounts for the two effects of the Coulomb
distortion mentioned above (momentum modification and focusing)
in the following way.
For a highly relativistic electron with zero impact parameter
the so-called effective momenta $k_{i,f}'$ of the electron are given by
\begin{equation}
k_i'=k_i+\Delta k, \quad k_f'=k_f+\Delta k, \quad k_{i,f}=|\vec{k}_{i,f}|,
\quad k_{i,f}'=|\vec{k}_{i,f}'|, \quad \Delta k = -V_0/c,
\end{equation}
where $V_0$ is the potential energy of the electron in the center
of the nucleus in analogy with eq. (\ref{parameters}).
E.g., for $^{208}$Pb we have $V_0 \sim -25\, \mbox{MeV}$,
not a negligible quantity when compared to energies of
some hundreds of MeV typically used in electron scattering experiments.
Cross sections are then calculated by using plane electron waves
corresponding to the effective momenta instead
of the asymptotic values in the matrix elements, and additionally
one accounts for the focusing factors of the incoming and
outgoing electron wave $k'_i/k_i$ and $k'_f/k_f$,
which both enter quadratically into the cross sections.
The main problem of the method is the fact that both
the focusing and the electron momentum are not constant inside
the nuclear volume. In the case of nucleon knockout reactions,
most of the hit nucleons are located near the surface of the nucleus,
where the classical momentum of the electrons is not given by the
central value.

A strategy to remedy this defect
is to alter the definition of the effective
momenta by not using the central potential value $V_0$,
but a value $V(r_f)$ obtained from some fitting procedure
(see also \cite{Traini2001} and references therein).
An even more ambitious strategy would be the introduction of
two effective momenta, one which accounts for the average modification of
the electron momentum inside the nucleus, and one which would
be utilized for the calculation of the average focusing of the electron wave
in the nuclear volume.

For quasielastic $(e,e')$ scattering a comparison
of EMA calculations with numerical results from the `exact'
DWBA calculation \cite{Onley92,Onley94} seems to indicate a failure of the EMA
\cite{JourdanWorkshop}. Also the improved approximation including
a first order correction from eq. (\ref{firstorder}) is of limited
validity. However, we find that the EMA with an effective
potential $\bar{V} \sim (0.75...0.8) V_0$ for heavier nuclei represents
a viable method for the analysis of Coulomb distortion effects
in inclusive quasielastic electron scattering, if the initial and final
energy of the electrons and the momentum transfer are sufficiently large.

\section{Exact numerical calculations}
\subsection{Properties of exact wave functions}
The nuclei of $^{40}$Ca and  $^{208}$Pb were chosen for our
calculations as typical examples for medium and highly charged
nuclei. The Dirac equation was solved by using a partial
wave expansion, which is discussed in detail in the appendix.
The radial integration of the radial functions was performed
by the method presented in \cite{Yennie54}. However,
we did not neglect the electron mass in our calculations,
although mass effects are quite small in our case.

We present results for an incoming electron
with spin in direction of the electron momentum
scattered off the fixed electrostatic potential of
the nucleus. Considering different spin or final
state waves with incoming spherical wave would lead basically to the same
conclusions.

The charge distribution of the $^{208}$Pb nucleus was modeled by a
Woods-Saxon distribution
\begin{equation}
\rho(r)=\frac{\rho_0}{e^{(r-r_{1/2})/a}+1}
\end{equation}
with $r_{1/2}=6.6$ fm and diffusivity $a=0.545$ fm,
compatible with an rms charge radius of $5.5$ fm and
a central Coulomb potential of $V_0=-25.7$ MeV,
whereas for the $^{40}$Ca nucleus
a three-parameter Fermi form was used
\begin{equation}
\rho(r)=\rho_0
\frac{1+\omega (r/r_{1/2})^2}{e^{(r-r_{1/2})/a}+1}
\end{equation}
with $r_{1/2}=3.766$ fm, diffusivity $a=0.586$ fm
and $\omega=-0.161$,
compatible with an rms charge radius of $3.48$ fm and
a central Coulomb potential of $V_0=-10.4$ MeV
\cite{Vries,Fricke}.

\begin{figure}
        \centering
        \includegraphics[width=9cm]{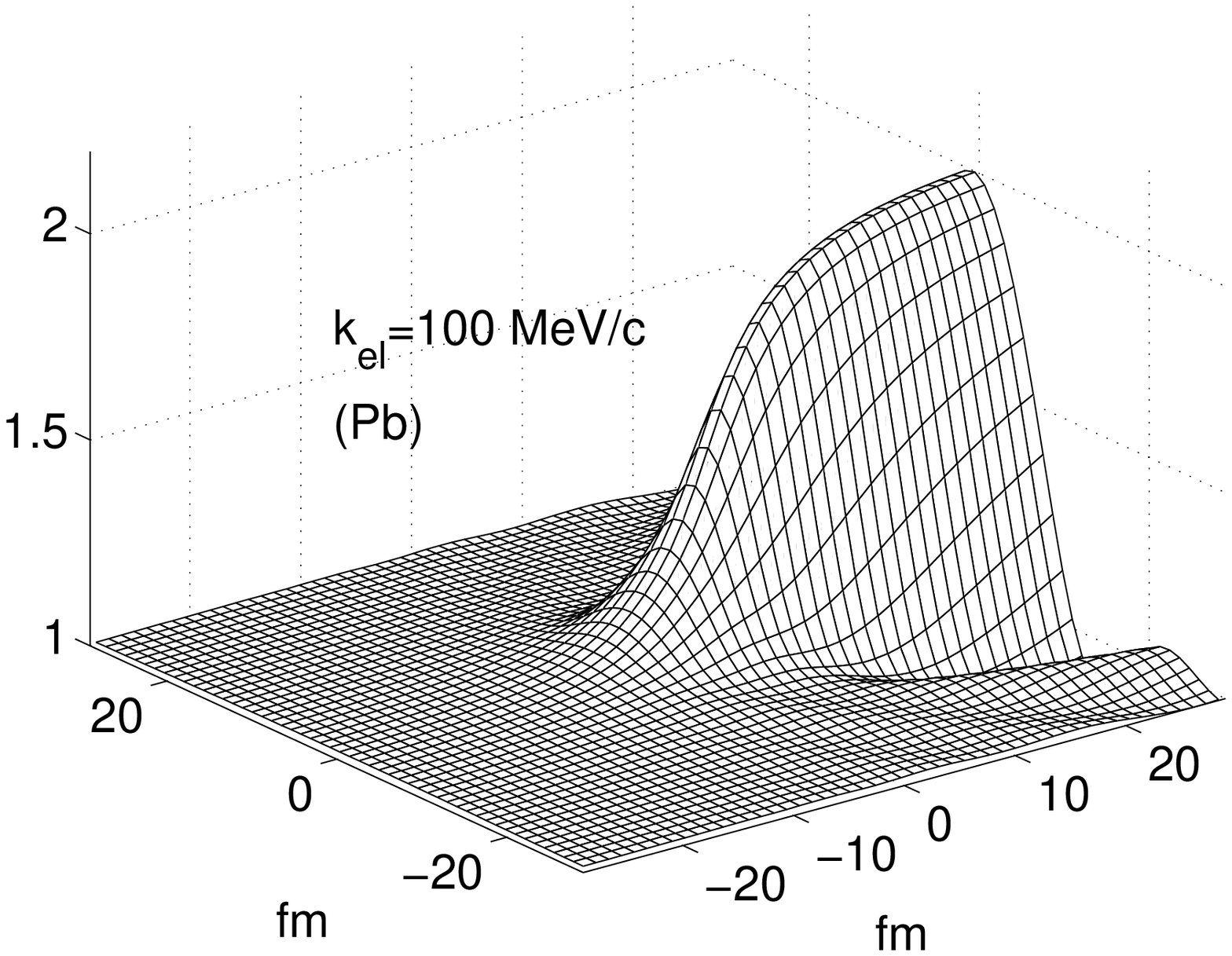}
        \caption{Amplitude $(\bar{\psi} \gamma^0 \psi)^{1/2}$}
                 of an electron wave incident on a $^{208}$Pb nucleus
                 with an energy of $100$ MeV. The focusing varies
                 strongly inside the nucleus, which has a radius
                 of the order of $7$ fm.
        \label{twodim}
\end{figure}

Fig. \ref{twodim} shows the focusing $(\bar{\psi} \gamma^0 \psi)^{1/2}$
of an electron wave incident
on a $^{208}$Pb nucleus with an energy of $100$ MeV.
The wave is normalized such that the density
$\bar{\psi} \gamma^0 \psi$ approaches the value $1$
in the asymptotic region.
The focusing is smaller in the upstream
side of the nucleus and grows larger in the downstream side.
At the same time, there is a strong decrease of the charge
density in transverse direction to the electron momentum.

\begin{figure}
        \centering
        \includegraphics[width=9cm]{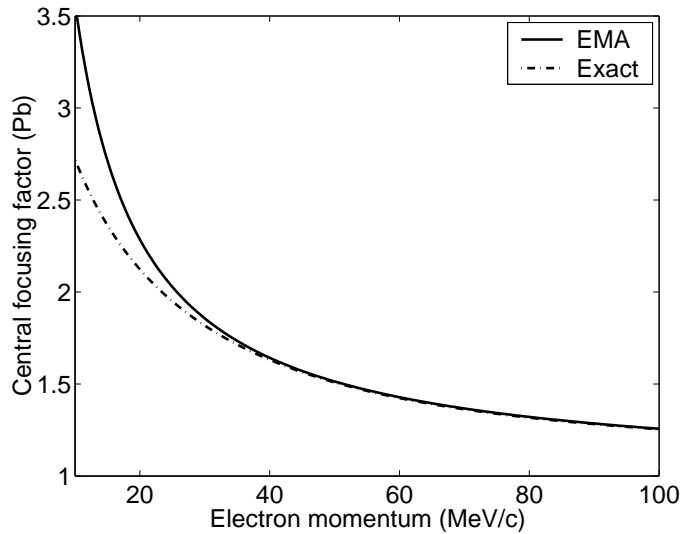}
        \caption{Focusing factor of the electron wave incident
                 on a $^{208}$Pb nucleus as a function of the
                 asymptotic electron momentum.}
        \label{central}
\end{figure}

As a first step, we checked the focusing factor in the
center ($r=0$) of the $^{208}$Pb nucleus. The exact central
focusing factor and the value typically used in the
EMA match extremely well already at 
relatively low energies above $40$ MeV, as shown in Fig.
\ref{central}.
For an electron energy of $100$ MeV the EMA focusing factor is
given by $(100+25.7)/100=1.257$,
as shown in Fig. (\ref{central}), and the exact value
deviates less than half a percent from this approximate result.
This positive result turned out to be generally valid for `well-behaved'
types of potentials like, e.g., Gaussian potentials with
depth and spatial extension comparable to
depth and extension of the nuclear potential.
But a closer look at the electron-wave amplitude
reveals that the amplitude varies strongly
inside the nuclear volume and the average focusing deviates
from the value calculated from the electrostatic potential $V_0$ in the
center of the nucleus.

\begin{figure}
        \centering
        \includegraphics[width=9cm]{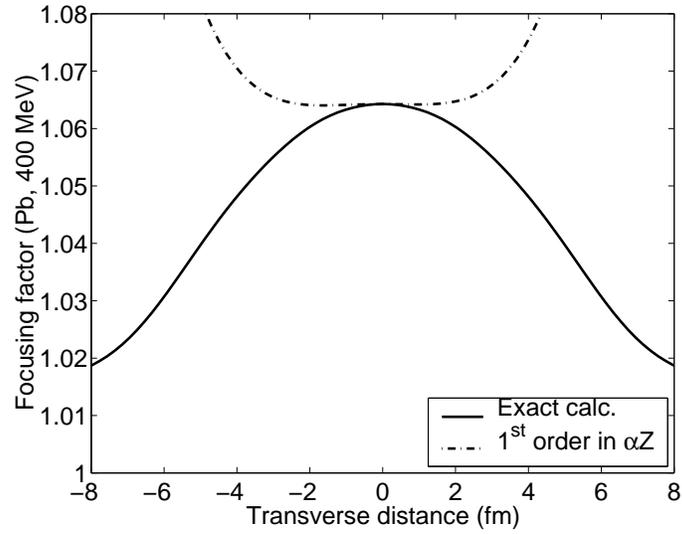}
        \caption{Focusing factor of the electron wave incident
                 on $^{208}$Pb with electron energy $400$ MeV,
                 on a straight line through the nuclear center
                 transverse to the electron momentum.}
        \label{transverse}
\end{figure}

Fig. \ref{transverse} shows the decay of the wave amplitude
on an axis perpendicular to the electron momentum which goes
through the nuclear center for an electron incident on
$^{208}$Pb with an energy of $400$ MeV. The central
focusing factor $\sim 425.7/400 \sim 1.0642$ decreases to
1.0187 at the outer edge of the nucleus at a transverse distance
of $8$ fm to the center.

\begin{figure}
        \centering
        \includegraphics[width=9cm]{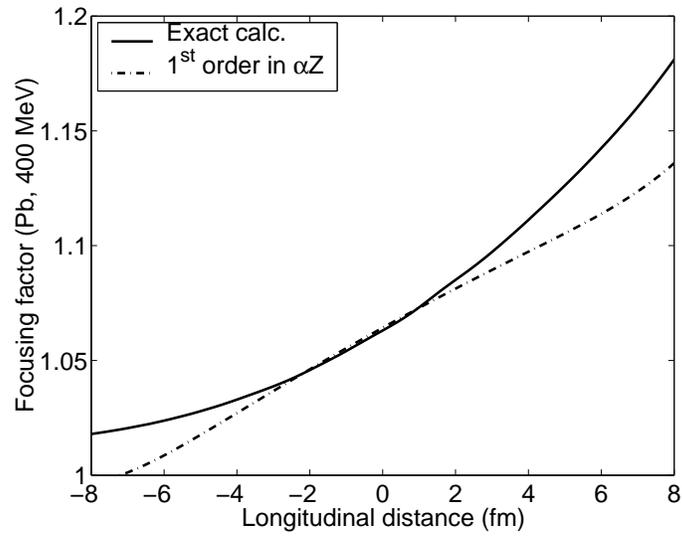}
        \caption{Focusing factor of the electron wave incident
                 on $^{208}$Pb with electron energy $400$ MeV,
                 along a straight line through the nuclear center
                 parallel to the electron momentum.}
        \label{longitudinal}
\end{figure}
Plotting the focusing factor along the axis through the
nuclear center parallel to the electron momentum shows
a strong increase on the downstream
side of the nucleus (Fig. \ref{longitudinal}).

One is therefore naturally lead to the idea to calculate
an averaged focusing factor $\bar{f}$ defined by
\begin{equation}
\bar{f}^{\, 2}=\frac{\int d^3 r \bar{\psi}_\tau (\vec{r})
\gamma^0 \psi_\tau (\vec{r}) \rho (r)}{\int d^3 r \rho(r)},
\label{focmean}
\end{equation}
where $\rho(r)$ is the nuclear matter density distribution
which can be well approximated by the charge density
profile of the nucleus for sufficiently large mass numbers
$A > 20$.
For a typical electron energy of $400$ MeV, one obtains
$\bar{f}=1.050$ for $^{208}$Pb,
corresponding to an effective potential
value of $-20.07$ MeV, in contrast to the often used potential depth
$V_0=-25.7$ MeV. The same calculation for
$^{40}$Ca leads to an effective potential of $-7.76$ MeV, compared
to $V_0=-10.4$ MeV.

Defining an effective potential value $\bar{V}$ by
\begin{equation}
\bar{V}=\frac{\int d^3 r \bar{\psi}_\tau (\vec{r})
\gamma^0 \psi_\tau (\vec{r}) \rho (r) V(r) }{\int d^3 r
\bar{\psi}_\tau (\vec{r}) \gamma^0 \psi_\tau (\vec{r}) \rho(r)},
\end{equation}
which is a measure for the average (semiclassical) electron momentum inside
the nuclear medium, leads to the very similar results
$\bar{V}=20.12$ MeV for $^{208}$Pb and $\bar{V}=7.78$ MeV
for $^{40}$Ca. The difference between the effective momenta
for the ingoing and outcoming electron calculated from this
effective potential value can also be viewed as the
effective momentum transferred by the electron to the
nucleon in a quasielastic knockout process.

These observations are a strong argument that
a modification of the EMA with an effective momentum
corresponding to an effective potential $\sim (0.75...0.8) V_0$
would provide reasonable results when experimental data
are corrected due to Coulomb distortion effects.
An effective potential value of $(18.7 \pm 1.5)$ MeV for $^{208}$Pb
was extracted by Gu\`eye et al. \cite{Gueye} by comparing
data from quasielastic positron scattering to quasielastic
electron scattering data taken by Zghiche et al. \cite{Zghiche}.
Exact calculations which include the Coulomb distortions
of positrons will be presented in a forthcoming paper.

It is interesting to note that also for lighter
nuclei like $^{40}$Ca a similar effective potential value
should be used like in the case of heavy nuclei.
But in such cases, Coulomb distortions are usually
of minor importance for the data analysis, and the
choice of the effective potential
value that is used in the EMA analysis plays a minor
role.

We further mention that the average potential
inside a homogeneously charged sphere is given by
$4 V_0/5$, where $V_0=-3 \alpha Z/2R$ is the value of the
potential in the center of the sphere. For positrons, we
found that the same effective potential (with opposite sign)
can be used, since the absolute values of the effective potentials
differ by less than $0.05$ MeV for an energy range of several
hundred MeV.

The first order term $g^{(1)}$
in eq. (\ref{expansionalpha})
fails to provide a satisfactory picture of the focusing inside the nucleus.
It does not reproduce the strong increase of the focusing on
the downstream side of nucleus, and the decrease of the focusing
in transverse direction is also not contained. On the contrary,
the dominant imaginary term $ib (\vec{k}' \times \vec{r} )^2$
causes an increase of the modulus of $1+g^{(1)}$ in transverse
direction. The first order term $g^{(1)}$ accounts for the
deformation of the wave front near the nuclear center,
but higher order terms are needed in order to
describe correctly the amplitude of the distorted
electron wave inside the nucleus. Calculations with a phenomenological
expression for the second order term $g^{(2)}$ have been presented
in \cite{Giusti88}.
The dash-dotted lines in Figs. \ref{transverse} and
\ref{longitudinal} show the focusing that would be obtained from
the first order expression $(k'/k)|1+g^{(1)}|$
for a homogeneously charged sphere with
a radius $R=7.1$ fm and $Z=82$, which is a good
approximation for a $^{208}$Pb nucleus.

It is solely the $a_1 r^2$-term which accounts for a decrease
of the focusing in transverse direction, but even if one neglects
the other terms in $g^{(1)}$ which cause an increase of
the focusing in transverse direction, the $a_1 r^2$-term
leads to a negligible effect compared to the actual
transverse decrease of the Coulomb distortion.
Therefore the assumption was made in \cite{Aste1,Aste2}
that the focusing is nearly constant in transverse direction;
the results presented there should be corrected
for the overestimated focusing. However, the results in
\cite{Aste1} are in good agreement with the exact calculations
presented by Kim et al. \cite{Jin1}, where exact
electron wave functions were used.

\begin{figure}
        \centering
        \includegraphics[width=9cm]{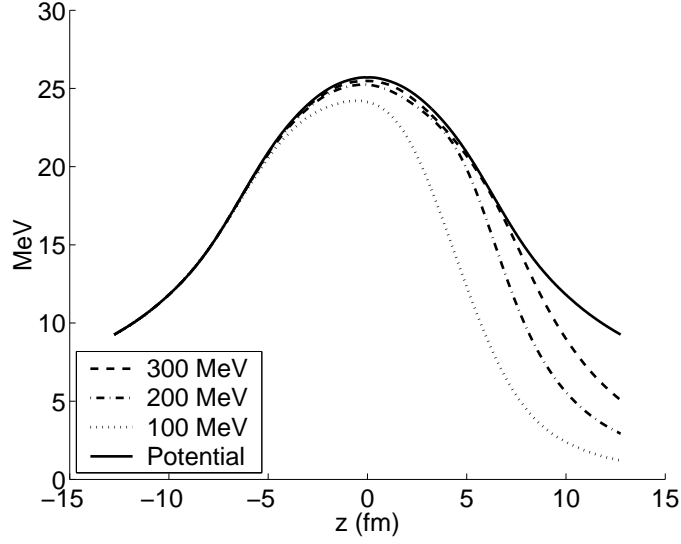}
        \caption{Comparison of (derivatives of) the phase of electron wave
                 functions along the $z-$axis for electrons incident on a
                 $^{208}$Pb nucleus with different energies.}
        \label{eikophase}
\end{figure}

A better approach than given by expansion
(\ref{expansionalpha}) to take the local change in
the momentum of the incoming particle into account is to modify
the plane wave describing the initial state of the particle
by the so-called eikonal phase $\chi_i(\vec{r})$ (see \cite{Aste1,Aste3} and
references therein)
\begin{equation}
e^{i\vec{k}_i\vec{r}} \, \rightarrow \,
e^{i \vec{k}_i\vec{r}+i\chi_i(\vec{r})} \, ,
\end{equation}
where
\begin{equation}
\chi_i(\vec{r})=-\int \limits_{-\infty}^{0} V(\vec{r}+
\hat{k}_i s) ds=
-\int \limits_{-\infty}^{z} V(x,y,z') dz' \, \label{int_eiko}
\end{equation}
if we set $\vec{k}_i=k_z^i \hat{{\bf{e}}}_z$.
In analogy to eq. (\ref{eiko1}), the $z$-component of the
momentum then becomes
in eikonal approximation
\begin{equation}
k_z e^{i k_z^i z+i\chi_i}=-i \partial_z e^{i k_z^i z+i\chi_i}=
(k_z^i-V)e^{i k_z^i z+i\chi_i}.
\end{equation}
The final state wave function is constructed analogously
\begin{equation}
e^{i \vec{k}_f\vec{r}-i\chi_f(\vec{r})},
\end{equation}
where
\begin{equation}
\chi_f(\vec{r})=-\int \limits_{0}^{\infty} V(\vec{r}+
\hat{k}_f s') ds' \, .
\end{equation}
In order to check the quality of this approximation,
we calculated the phase
of the first (large) component $\psi^1_{1/2}$
from the exact electron spinor along the $z$-axis, and extracted
the quantity $\chi^{ex}_i$ by setting
\begin{equation}
e^{i k_i z+i\chi^{ex}_i(z)}=\psi^1_{1/2}(z)/|\psi^1_{1/2}(z)|.
\end{equation}
If the eikonal approximation (\ref{int_eiko}) were exact, then the
derivative $\frac{d}{dz} \chi^{ex}_i(z)$
would be equal to the negative value of the electrostatic potential
\begin{equation}
\frac{d}{dz} \chi^{ex}_i(z) = -V(z).
\end{equation}
In Fig. (\ref{eikophase}) a comparison of $\frac{d}{dz} \chi^{ex}_i(z)$
to the absolute value of the electrostatic potential of $^{208}$Pb is
shown. For energies above $300$ MeV, the phase of the electron wave function
is very well described by eq. (\ref{int_eiko}) inside the nuclear volume.

The validity of our calculations was verified by reinserting the
electron wave functions into the Dirac equation. This way, the
electrostatic potential can be reproduced and compared to the initial
potential. Additionally, we checked current conservation, the
asymptotic behavior of the wave functions far away from the nucleus and
the behavior for the limit $Z \rightarrow 0$.
Finally, the calculations were also performed
for the potential of a homogeneously charged sphere. In this case,
analytic expressions are available for the radial wave functions occurring
in the partial wave expansion \cite{Pauli}, which were in perfect
agreement with the results obtained via radial integration. 

\subsection{EMA from DWBA}
We establish now a connection between the
DWBA results and the EMA. The DWBA transition amplitude for
high momentum transfer in inelastic electron scattering
can be written for one-photon exchange as
\begin{equation}
T_{if}=\int d^3 r_e d^3 r_N \Bigl\{ \rho_e (\vec{r}_e) \rho_{if} (\vec{r}_N)-
\vec{j}_e(\vec{r}_e) \vec{J}_{if}(\vec{r}_N) \Bigr\}
\frac{e^{i \omega r}}{|\vec{r}_e-\vec{r}_N|}
\end{equation}
with $\rho_e, \, \rho_{if}, \, \vec{j}_e, \, \vec{J}_{if}$ being the charge
and current densities of the electron and the nucleus, respectively,
and $\omega$ is the energy loss of the electron. The double volume integral
presents a clear numerical disadvantage of this expression.
According to Knoll \cite{Knoll1}, one may introduce the scalar operator
\begin{equation}
S=e^{i \vec{q}\vec{r}} \sum \limits_{n=0} \Biggl(
\frac{2i \vec{q} \, \vec{\nabla} +\Delta}{\vec{q}^{\, 2}-\omega^2} \Biggr)^n
e^{-i \vec{q} \vec{r}}, \quad \vec{q}=\vec{k}_i-\vec{k}_f, \label{scalarop}
\end{equation}
such that the transition amplitude can be expanded in a more convenient
form ($Q^2=\vec{q}^{\, 2}-\omega^2$)
\begin{equation}
T_{if}= \frac{4 \pi}{Q^2} \int d^3 r
\Bigl[ \rho_{if} (\vec{r}) S \rho_e (\vec{r}) - \vec{J}_{if} S \vec{j}_e (\vec{r}) \Bigr].
\end{equation}
The single integral is limited to the region of the nucleus, where the nuclear current
is relevant. The expansion (\ref{scalarop}) is an asymptotic one, which means that there
is an optimum number (depending on $\vec{q}$) of terms that give the best approximation
to the exact value.
Considering terms up to second order in the derivatives only one obtains
\begin{equation}
T_{if} =\frac{4 \pi}{Q^2} \int d^3 r \Biggl\{
\rho_{if} (\vec{r}) e^{i \vec{q} \vec{r}} \Biggl[
1 + \frac{2 i \vec{q} \, \vec{\nabla} + \Delta}{Q^2}
-\frac{4 ( \vec{q} \,  \vec{\nabla})^2}{(Q^2)^2} \Biggr]
e^{-i \vec{q} \vec{r}} \rho_e(\vec{r}) + \mbox{current terms} \Biggr\}. \label{approx}
\end{equation}
Note that the $e^{-i \vec{q} \vec{r}}$-term in front of the electron charge
or current density cancels the spatial oscillatory behavior
of the densities in the asymptotic region far from the nucleus, such that
the gradient operator $\vec{\nabla}$ in eq. (\ref{approx}) probes only
the distortion of the electron current due to the nuclear Coulomb potential.
Therefore, eq. (\ref{approx}) provides a satisfactory approximation for many
interesting cases when $Q^2 \gg q \Delta k$ is fulfilled. E.g., for a typical
initial electron energy $\epsilon_i=400$ MeV used in inclusive quasielastic
electron scattering experiments and energy transfer
$\omega=\epsilon_i-\epsilon_f=100$ MeV,
the contribution to the cross section due to the
$(\vec{q} \, \vec{\nabla})^2$-term in eq. (\ref{approx}) is only of the
order of 2\% for momentum transfer $q \ge 350$ MeV \cite{Jin1,Aste1}.

Expansion (\ref{approx}) allows for a clear comparison of the plane wave
Born approximation, the EMA and the DWBA. The effect of the electrostatic nuclear
field on the matrix element $T_{if}$ is obviously given by a modification of the
free electron charge and current densities $\rho_e$,
$\vec{j}_e$ via the replacements
\begin{equation}
[\rho_e(\vec{r}),\vec{j}_e (\vec{r})] \rightarrow
[\rho_e' (\vec{r}),\vec{j}_e' (\vec{r})]=
e^{i \vec{q} \vec{r}} 
\Biggl[1 + \frac{2 i \vec{q} \, \vec{\nabla} + \Delta}{Q^2}
-\frac{4 ( \vec{q} \, \vec{\nabla})^2}{(Q^2)^2} \Biggr]
e^{-i \vec{q} \vec{r}} [\rho_e(\vec{r}),\vec{j}_e (\vec{r})].
\end{equation}
Having exact wave functions at hand, it is possible to perform
a numerical analysis of this modification.
For this purpose, it is advantageous to introduce the effective
four-momentum transfer squared given by
$Q'^2=(\vec{k}_i'-\vec{k}_f')^2-\omega^2$.
Then a short calculation shows that ($\epsilon_{i,f} \gg m$)
\begin{equation}
\frac{Q'^2}{Q^2}=\frac{k_i' k_f'}{k_i k_f}, \label{ratioQ}
\end{equation}
i.e., when cross sections are calculated using the EMA,
the enhanced photon propagator appearing in the matrix element
$T_{if}$ exactly cancels the focusing effect of the initial and
final state wave function.
A numerical analysis shows that this is indeed also true
for the DWBA case to a high degree of accuracy.
E.g., a calculation for relatively low electron energies
$\epsilon_i=300$ MeV, $\epsilon_f=200$ MeV and scattering angle
$\vartheta_e=60^o$ of
\begin{equation}
\bar{f'}^2=\frac{\int d^3 r | \rho_e'(\vec{r}) | \rho(\vec{r})}
{\int d^3 r \rho(\vec{r})}, \quad
\bar{f}^2_{\mbox{\footnotesize{free}}}=\frac{\int d^3 r
| \rho_e^{\mbox{\footnotesize{free}}}(\vec{r}) | \rho(\vec{r})}
{\int d^3 r \rho(\vec{r})}
\end{equation}
in analogy to eq. (\ref{focmean}) leads for $^{208}$Pb to
the numerical result $\bar{f'}^2/\bar{f}^2_{\mbox{\footnotesize{free}}}
=0.985$, i.e. the wave function focusing effect of $\sim 17\%$
in the electron charge density is in fact
overcompensated slightly by $\sim 1.5\%$ compared to the free
plane wave charge density due to the enhanced
momentum transfer. The situation is similar for a backscattering angle
$\vartheta_e=143^o$, which has also been used in the experiment
\cite{Zghiche}.
Numerical calculations reveal that the same observation
applies to the components of the current density.
Defining focusing factors $\bar{f'}^2_k$ for each current component
$\vec{j}_e=(j_x,j_y,j_z)$ again leads to the result that
the effect of the $S$-operator is just a cancellation of the average
focusing effect in the initial and final state wave functions.

Finally, we investigated the modification of the momentum transfer
inside the nucleus. For this purpose, we defined the local momentum
transfer $\vec{q}(\vec{r})$ according to
\begin{equation}
\rho_e'(\vec{r})=|\rho_e'(\vec{r})| e^{i \vec{q}(\vec{r}) \vec{r}}
\label{rhodecomp}
\end{equation}
and calculated the quantities
\begin{equation}
q_1'=\frac{\int d^3 r |\vec{q}(\vec{r})| \rho(\vec{r})}{\int d^3 r \rho(\vec{r})},
\quad
{q'}_{\, 2}^2=\frac{\int d^3 r |\vec{q}(\vec{r})|^2 \rho(\vec{r})}{\int d^3 r \rho(\vec{r})}
\end{equation}
The results for $q_1'$ and $q_2'$ are again compatible with an average potential
of $20$ MeV. Replacing $\rho_e'$ by $\rho_e$ in eq. (\ref{rhodecomp})
has nearly no influence on the result for the corresponding effective momenta
$q_1$ and $q2$, and again the situation is completely analogous for
the current density.

\subsection{Coulomb corrections for a harmonic oscillator single particle shell model}
We point out that the discussion presented above is a most general one
as long as detailed properties of the nuclear current are ignored.
The hope is of course that the presence of, e.g., 208 nucleons
in a $^{208}$Pb nucleus leads to a smooth nucleon momentum distribution,
such that the features of the individual nucleon wave functions are averaged out
to some degree in the momentum regime relevant for our discussion.
Calculations in the framework of the eikonal approximation as presented in \cite{Aste1},
but with the electron amplitudes adapted to the amplitudes from exact calculations,
also agree well with the EMA, a fact which supports the assumption that a
semiclassical description of Coulomb corrections can be found at high momentum transfer.
Still, one cannot exclude from the first that the semiclassical picture
of the transition amplitude which has been presented in the previous
discussion is flawed by phase effects which enter the cross section, leading
to unexpected deviations from the idealized EMA-behavior.

We therefore present results for a Pb model nucleus
consisting of 82 scalar protons with harmonic nucleon wave functions
for some typical kinematical situations, which indeed show
that the modified EMA, using an average potential value,
reproduces the Coulomb effects with high accuracy for
$(e,e')$ scattering, if the momentum transfer and the energy of the
outgoing electron are high enough. The smaller contribution of the neutrons
was neglected in the calculations, since we will focus on general considerations in the
following. A detailed study using relativistic nucleon wave functions including spin
in conjunction with more realistic current models will be presented in detail in a
forthcoming paper.

At the lowest energies that have been used in the positron experiment
by Gu\`eye et al. \cite{Gueye}, i.e., for an initial electron energy
$\epsilon_i^-=224$ MeV and initial positron energy $\epsilon_i^+=262$ MeV,
the situation is a bit more involved, and we found a relatively
large disagreement between the exact inclusive cross sections and
the values calculated from the EMA with an effective potential value
of $19$ MeV, as will be demonstrated below. 
This observation is not astonishing, since the typical energy of the outgoing electrons
is only of the order of $100$ MeV in this case (see also Fig. 5).

The radial wave function $R_{nl}$ of a particle in the harmonic oscillator
shell model potential $V(r)=-v_0+ \frac{1}{2} M \omega_o^2 r^2$, where
$M$ is the proton mass, is given by ($n=1,2...$ and $l=0,1...)$
\begin{equation}
R_{nl}(r)=\frac{1}{\sqrt{x_o^3}}N_{nl} \Biggl( \frac{r^l}{x_o^l}
\Biggr) L_{n-1}^{l+\frac{1}{2}} \Biggl( \frac{r^2}{x_o^2} \Biggr)
e^{-\frac{1}{2} \frac{r^2}{x_o^2}}, \quad x_o^2=\frac{\hbar}{M \omega_o}
\end{equation}
such that $R_{nl}$ fulfills the normalization condition
$ \int_0^{\infty} R_{nl}^2(r) \, r^2 \, dr = 1$
and the integral $\langle n \, l | r^2 | n \, l \rangle$ has the value
$(N=2n+l-2)$
\begin{equation}
\langle r^2 \rangle = \int_0^{\infty} R_{nl}^2(r) \, r^4 \, dr =
x_o^2 (N+\frac{3}{2}), \label{staterms}
\end{equation}
with normalization constants
\begin{equation}
N_{nl}^2=\frac{2^{n+l+1}}{\sqrt{\pi}(n-1)!(2n+2l-1)!!}.
\end{equation}
The Legendre polynomials are given by
\begin{equation}
L_{n-1}^{l+\frac{1}{2}}(z)=\sum_{k=0}^{n-1} \frac{(-1)^k}{2^{n-k-1}}
{n-1 \choose k} \frac{(2n+2l-1)!!}{(2l+2k+1)!!} z^k
\end{equation}
and the energies are $E_N=-v_0+\hbar \omega_o (N+3/2)$.
A popular phenomenological choice for the oscillator strength is $\hbar \omega_o
\sim 41 \, \mbox{MeV} \, A^{-1/3}$.
The $(n,l)=(1,5)$-shell corresponds to the
$1h_{11/2}$-shell in the spin-orbit interaction model and contains
12 protons (but $2(2l+1)=22$ available states, when an artificial
spin factor of $2$ is included), whereas the the $(n,l)=
(1,0)$, $(2,0$), $(3,0)$, $(1,1)$, $(1,2)$, $(2,1)$, $(2,2)$, $(1,3)$, and
$(1,4)$-shell are fully occupied.
We therefore assumed for the calculation of cross sections that the
$(1,5)$-shell is completely filled, i.e. spherically symmetric, and adopted
a weighting factor of $12/22$ leading to an rms charge radius of the Pb
model nucleus of $r_{_{rms}}=\sqrt{393/82} x_0$, as can be derived
from eq. (\ref{staterms}). Accordingly, $x_o=2.515$ fm was used
in the calculations, such that the rms charge radius for the nucleus adds up to $5.505$ fm.

The proton transition current density was calculated from the free form of the Klein-Gordon
expression
\begin{equation}
\vec{j}_p=\frac{i e}{2 M} \Bigl\{ \Phi^*_f \vec{\nabla} \Phi_{nlm} -
\Phi_{nlm} \vec{\nabla} \Phi^*_f \Bigr\} \label{protoncurrent}
\end{equation}
for initial states $\Phi_{nlm}(\vec{r})=R_{nl}(r) Y_{lm}(\vec{r}/r)$
and final plane wave states $\Phi_f=e^{i \vec{k}_f^p \vec{r}}$.
The transition charge density was calculated from the exact current conservation relation
\begin{equation}
i \omega j_p^0=\vec{\nabla} \vec{j}_p, \label{zerocomp}
\end{equation}
where $\omega=\epsilon_i-\epsilon_f$ is the energy transferred to the proton
as defined before. The model used for the proton current
is simple, but it has the advantage that it allows to perform analytical calculations
for the electron plane wave case, which permit a verification of the accuracy of the numerical
calculations. It also captures the most important features of the $(e,e')$ cross
sections in the vicinity of the quasielastic peak. Furthermore, it has to be pointed
out that the Ohio group used a single particle shell model, where the nucleon wave
functions were obtained by solving the Dirac equation for each shell nucleon with
phenomenological S-V-potentials. But a comparison
of measured data and calculations shows large discrepancies especially at higher
energy transfer $\omega \ge 150$ MeV, where correlation effects and pion
production become increasingly important. Therefore also a single particle
shell model with `exact' nucleon wave functions {\emph{cannot}} be considered as an
`exact' model for inclusive quasielastic scattering.
Hence, the {\emph{optimal strategy}} is to find a general, model independent
method which makes it possible to include the Coulomb distortion effect in
the analysis of experimental data. 

For lower electron energies, where the EMA cross sections
($\sigma_{_{EMA}}$) start to deviate significantly from the exact ones ($\sigma_{_{CC}}$),
it is still possible to match the EMA and exact cross sections
by using an effective potential value $\bar{V}_{fit}$ that differs
from the commonly used $18-20$ MeV for Pb.
E.g., for $\epsilon_i^-=224$ MeV, $\sigma_{_{CC}}$ and $\sigma_{_{EMA}}$ are very close
for $\bar{V}_{fit} \sim 27$ MeV in our model for an energy transfer larger than $100$ MeV.
This effective value is very stable
under distortions of the scalar proton current model; using a `wrong' value for
$\omega$ in eq. (\ref{zerocomp}) which is $20$ MeV larger or smaller than
$\epsilon_i-\epsilon_f$ or including a potential term of similar order in
eq. (\ref{protoncurrent}) changes the cross sections, but the effective potential value
does not change significantly. This is a positive result, since it shows that our general
considerations are not too strongly model dependent. Binding energies, nuclear potentials and the
corresponding exact nucleon wave functions are relevant for an accurate modelling of
amplitude, width and position of the quasielastic peak, but the impact of
the Coulomb distortion of electrons on the quasielastic cross section expressed by
a fitted effective potential $\bar{V}_{fit}$
can also be studied to some level using a simplified model with a quasielastic
peak that incorporates approximately the properties of the true quasielastic
cross section. Since binding energies are indeed neglected in our theoretical model,
a shift of the quasielastic peak by about $25$ MeV to lower energy transfer is observed.

It is very instructive to investigate the total response function $S^{tot}$,
which is defined in plane wave Born approximation by
\begin{equation}
\frac{d^2 \sigma_{_{PWBA}}}{ d \Omega_f d\epsilon_{f}}=
\sigma_{Mott} \times S^{tot}_{_{PWBA}}
(|\vec{q} \, |,\omega,\Theta),
\end{equation}
via the well-known Mott cross section
\begin{equation}
\sigma_{Mott}=4 \alpha^2 \cos^2(\Theta/2) \epsilon_f^2/Q^4
\end{equation}
not only for the total cross section, but also for individual filled shells.
According to eq. (\ref{ratioQ}), the Mott cross section remains unchanged
when it gets multiplied by the EMA focusing factors
and the momentum transfer $Q^4$ is replaced by its
corresponding effective value. Therefore, if the EMA
is a good approximation, the cross section can be written (see also \cite{Gueye})
\begin{equation}
\frac{d^2 \sigma_{_{CC}}}{ d \Omega_f d\epsilon_{f}} \simeq
\frac{d^2 \sigma_{_{EMA}}}{ d \Omega_f d\epsilon_{f}}=
\sigma_{Mott} \times S^{tot}_{_{PWBA}}(|\vec{q}_{_{eff}}|,\omega,\Theta).
\end{equation}

Table 1 shows ratios of total responses, which have been obtained
by dividing the same Mott cross section out of the inclusive Coulomb corrected,
EMA-, and PWBA-cross sections restricted to single closed shells. The first two
columns show a case where both the initial and final electron energy is
larger than $300 $ MeV, the other four columns show two kinematical settings
where both the initial and final electron (positron) energy is
smaller than $300$ MeV. In all three cases, an effective potential value
of $19$ MeV was used.

It is an interesting point is that even closed shells show an EMA-like behavior
at higher electron energy. This is indeed not the case for
single states, which are not spherically symmetric. Summing over all shells,
one obtains a ratio $S^{tot}_{_{CC}}/S^{tot}_{_{EMA}}$ which is practically one
for an effective potential value $\bar{V}=19$ MeV in the case where $\epsilon_i=485$ MeV.
Note that the different shells contribute differently to the total cross section,
according to the number of protons contained in them and their momentum distribution.
E.g., the (1,0)-shell, which show the most irregular behavior for the kinematics
shown in Table 1 due to its narrow momentum distribution, contributes only marginally
to the total cross section because the shell contains only 2 protons.

In the case given in Table 1 with an initial electron energy $\epsilon_i^-=224$ MeV,
the exact cross section is $7.3\%$ larger than the EMA cross section and $24.4\%$ larger
than the plane wave cross section. By naive interpolation, one can infer that
the exact and the EMA cross section should be identical for an effective potential
value which is larger than $19$ MeV. This indeed the case for an
approximate value $\bar{V}_{fit} \simeq 27$ MeV, and similarly in the positron case
for $\bar{V}_{fit} \simeq 25$ MeV. The maximum values of the total response for
the $\epsilon_i^-=224$ MeV and $\epsilon_i^+=262$ MeV case differ by about $10\%$,
i.e., one observes a relatively large deviation from the EMA prediction that the
two responses agree. However, for higher energies used in the
positron experiment of Gu\`eye et al., an effective potential
value of $18.9 \pm 1.5$ MeV {\emph{is}} compatible with our model calculations.

The accuracy concerning the calculation of cross sections and accordingly
the values given in Table 1 is limited to about $1\%$ due to the truncation of the Knoll
expansion eq. (\ref{approx}) and the finite resolution of the grid that
has been used for the modelling of the nucleus. 
The numerical evaluation of transition amplitudes was performed by putting
the nucleus on a three dimensional cubic grid with a side length of $30$ fm and
a grid spacing of $(30/140)$ fm, and convergence was checked by using different side lengths
and grid resolutions. The accuracy of the solid angle integration of the
$(e,e'p)$ cross section was better than $0.05\%$.

\begin{table}
\begin{center}
\begin{tabular}{|c|cc|cc|cc|}
	\hline
 & & & & & & \\
$(n,l)$                                           &
$S^{tot,nl}_{_{CC}}/S^{tot,nl}_{_{\! EMA}}$       &
$\! S^{tot,nl}_{_{CC}}/S^{tot,nl}_{_{\! PWBA}}$   &
$S^{tot,nl}_{_{CC}}/S^{tot,nl}_{_{\! EMA}}$       &
$\! S^{tot,nl}_{_{CC}}/S^{tot,nl}_{_{\! PWBA}}$   &
$S^{tot,nl}_{_{CC}}/S^{tot,nl}_{_{\! EMA}}$       &
$\! S^{tot,nl}_{_{CC}}/S^{tot,nl}_{_{\! PWBA}}$ \\
 & & & & & & \\
	\hline
 & & & & & & \\
(1,0)   & 1.101 & 1.811   &   1.275 & 3.574   &   0.776 & 0.425 \\
(2,0)   & 1.004 & 0.971   &   0.996 & 0.972   &   1.056 & 0.679 \\
(3,0)   & 1.013 & 1.046   &   1.116 & 1.272   &   1.033 & 0.870 \\
(1,1)   & 1.024 & 1.332   &   1.140 & 2.041   &   0.896 & 0.758 \\
(1,2)   & 0.989 & 1.094   &   1.063 & 1.384   &   0.964 & 0.982 \\
(2,1)   & 1.013 & 1.268   &   1.050 & 1.377   &   1.025 & 0.718 \\
(2,2)   & 0.995 & 1.271   &   1.031 & 1.526   &   1.028 & 0.939 \\
(1,3)   & 0.982 & 0.989   &   1.054 & 1.132   &   1.003 & 1.087 \\
(1,4)   & 0.990 & 0.956   &   1.072 & 1.043   &   1.011 & 1.115 \\
(1,5)   & 1.005 & 0.955   &   1.085 & 1.011   &   1.006 & 1.117 \\ 
 & & & & & & \\
	\hline
 & & & & & & \\
$\sum \limits_{\, (n,l)'}$  & 0.999 & 1.072 & 1.073 & 1.244 & 0.989 & 0.956 \\
	\hline
\end{tabular}
\end{center}
\caption{Ratio of the total response for different filled shells, calculated
in plane wave approximation ($S^{tot}_{_{PWBA}}$) and EMA with an effective
potential $19$ MeV ($S^{tot}_{_{EMA}}$), with the exact response $S^{tot}_{_{CC}}$.
First two columns: Initial electron energy $\epsilon_i=485$ MeV, $\omega=140$ MeV,
$\vartheta_e=60^o$. Columns 3 \& 4: Initial electron energy $\epsilon_i=224$ MeV,
$\omega=100$ MeV, $\vartheta_e=143^o$. Columns 5 \& 6: Initial positron energy
$\epsilon_i=262$ MeV, $\omega=100$ MeV, $\vartheta_p=143^o$. The bottom line shows
the ratios of the total response functions for the full cross sections, i.e. after
summing over all shells with a weighting factor 6/11 for the (1,5)-shell.
Note that these values are not averages of the data in the
column above, since the contribution of each shell to the total cross section is different.}
\end{table}

\section{Conclusions}
The complex behavior of the Coulomb distortion of electron waves
at relatively high energies relevant for quasielastic
electron-nucleus scattering experiments was studied using
accurate numerical calculations. Naive lowest order approximations in
$\alpha Z$ are not suitable for the analysis of Coulomb
corrections in scattering experiments, unless they
are modified in a well-controlled manner based on exact
calculations.
A Fortran 90 program is available now which can be
used for accurate calculations of continuum electron wave
functions in a central electrostatic field. The exact
wave functions were used for a numerical study of Coulomb distortions
in inclusive quasielastic electron scattering. It turns out that
the effective momentum approximation is not
reliable when the central potential value $V_0$ of the
electrostatic field of the nucleus is taken as basis for
the EMA calculations, but a smaller average value
$\bar{V} \sim (0.75...0.8) V_0$ leads to very good results, if the momentum transfer
and the energy of the scattered electron large enough.
An effective potential value of $19$ MeV is a very good choice for
$\epsilon_f \geq 300$ MeV and $Q^2 \geq (400 \, \mbox{MeV})^2$,
and we conjecture that at very high energies, a limiting effective potential
value close to $20$ MeV is reached.
If the energy of the scattered electron becomes smaller than $300 $ MeV,
the semiclassical description of the final state wave function
becomes obsolete, but it is still possible to use an EMA-like approach
for the description of the inclusive cross section by using a modified fitted
potential value, given the condition that the initial and final energy of the
electron and the momentum transfer are not too small. In this region,
detailed calculations become necessary with more refined nuclear models
that the one used in this work. However, for the kinematical settings
that will be used in the future experiments at the TJNAF, our analysis
shows that the EMA will provide a valuable strategy for the correction
of data.

It highly improbable that the match of our exact
cross sections with the EMA cross sections (i.e., with $\bar{V}=20$ MeV
for $^{208}$Pb), which is better than $2\%$ for the kinematical region
 $\epsilon_f \geq 300$ MeV and $Q^2 \geq (400 \, \mbox{MeV})^2$, is just a pure
coincidence, since both approaches are unrelated and based on different
calculational strategies.
We must therefore conclude that our findings are not
compatible with the conclusions drawn in \cite{Jin1}.

\section*{Acknowledgements}
The authors would like to thank Zein-Eddine Meziani, Joseph Morgenstern,
Paul Gu\`eye, John Tjon, Jian-Ping Chen, Seonho Choi,
Ingo Sick, J\"urg Jourdan, and Kai Hencken
for interesting and useful discussions. Part of this work was presented
at the {\emph{Mini-Workshop on Coulomb Corrections}}, Thomas 
Jefferson National Accelerator Facility, March 2005, and at
the Joint Jefferson Lab/Institute of Nuclear Theory Workshop on
Precision ElectroWeak Interactions, College of William and Mary, August 2005.
This work was supported by the Swiss National Science Foundation.

\section*{Appendix: Calculation of continuum states of an
electron in a central electrostatic field}
Some details concerning the description of electron
continuum states in a central electrostatic field are given here
for the reader's convenience and in order to give a
fully consistent description of the problem.

We first consider the solutions of the 
stationary Dirac equation
\begin{equation}
[-i \vec{\alpha} \vec{\nabla}+ m \beta +V(\vec{r})] \psi(\vec{r})=
E \psi(\vec{r})
\end{equation}
for an electron with total energy $E$
subject to the central electrostatic potential generated by a spherically
symmetric nucleus with charge number $Z$.
We use standard Dirac and Pauli matrices \cite{Bjorken}.
Dirac spinors describing states with definite angular
momentum and parity can be decomposed into a radial and an angular
part
\begin{equation}
\psi_{\kappa}^{ \mu} = \left(
\begin{array}{c}
g_{\kappa}(r)\chi_{\kappa}^{\mu} (\hat{\bf{r}}) \\
i f_{\kappa}(r) \chi_{-\kappa}^{\mu} (\hat{\bf{r}})
\end{array} \right) \label{sphericalspinor}
\end{equation}
with two-component spinors $\chi_{\kappa}^{\mu}$ which are
eigenstates of the spin-orbit operator
\begin{equation}
(\vec{\sigma} \vec{L} +1) \chi_{\kappa}^{\mu}=
-\kappa \chi_{\kappa}^{\mu}
\end{equation}
and the angular momentum operators
\begin{displaymath}
\vec{J}^{\, 2}  \chi_{\kappa}^{\mu}=
(\vec{L}+\vec{s} \, )^{2}  \chi_{\kappa}^{\mu}=j(j+1) \chi_{\kappa}^{\mu},
\end{displaymath}
\begin{displaymath}
\vec{L}^2  \chi_{\kappa}^{\mu}=l(l+1) \chi_{\kappa}^{\mu} \, , \quad
\vec{s}^{ \, 2}  \chi_{\kappa}^{\mu}=\frac{3}{4} \chi_{\kappa}^{\mu},
\end{displaymath}
\begin{equation}
J_z \chi_{\kappa}^{\mu}  = \mu \chi_{\kappa}^{\mu} \, ,
\end{equation}
where $\kappa = \pm 1,\pm 2,...$ is related to $j$ and $l$ by
\begin{equation}
\kappa=l(l+1)-(j+\frac{1}{2})^2 , \quad
j=|\kappa|-\frac{1}{2}  , \quad
l=j+\frac{1}{2} \mbox{sgn} (\kappa) ,
\end{equation}
and the operators $\vec{L}$ and $\vec{s}$ are given by
$\vec{L}=-i \vec{r} \times \vec{\nabla}$ and
$\vec{s}=\frac{1}{2} \vec{\sigma}$.
The spinors can be expressed using Clebsch-Gordan
coefficients as
\begin{equation}
\chi_{\kappa}^{\mu}  =  \sum_{\zeta=\pm 1/2}
\langle l (\mu-\zeta) \frac{1}{2} \zeta | j \mu \rangle
Y_l^{\mu-\zeta} ({\bf \hat r}) \chi_{\zeta}, \quad
\hat j  =  \sqrt{2j+1}
\end{equation}
where $\chi_{\zeta}$ are standard Pauli spinors, or more explicitly
\begin{equation}
\chi_{\kappa}^{\mu}  = 
\left( \begin{array}{c} 
\sqrt{\frac{j+\mu}{2 j}} Y_{l,\mu-\frac{1}{2}} \\
\sqrt{\frac{j-\mu}{2 j}} Y_{l,\mu+\frac{1}{2}}
\end{array} \right) \, \, \mbox{for} \, \, \kappa < 0 , \quad
\chi_{\kappa}^{\mu}  = 
\left( \begin{array}{c} 
-\sqrt{\frac{j-\mu+1}{2 j+2}} Y_{l,\mu-\frac{1}{2}} \\
\sqrt{\frac{j+\mu+1}{2 j+2}} Y_{l,\mu+\frac{1}{2}}
\end{array} \right) \, \, \mbox{for} \, \, \kappa > 0.
\end{equation}

The radial functions $f_{\kappa}$ and $g_{\kappa}$
fulfill the coupled differential equations
\begin{equation}
\frac{d}{dr} \left( \begin{array}{c} 
g_{\kappa}\\
f_{\kappa} \end{array} \right) = \left( \begin{array}{cc}
-\frac{\kappa+1}{r} & E+m-V \\
-(E-m-V) & \frac{\kappa -1}{r} \end{array} \right)
\left( \begin{array}{c}
g_\kappa \\ f_\kappa \end{array} \right) \, . \label{cdg}
\end{equation}
For an electron with energy $E>0$ in the Coulomb field of a
point like charge $eZ$ the potential $V$ is
\begin{equation}
V(r) = -\xi/r, \quad  \xi=\alpha Z,
\end{equation}
and the continuum solutions of the Dirac equation are given by (see
\cite{TBR83} and references therein)
\begin{eqnarray}
\left( \begin{array}{c} 
g_\kappa \\
f_\kappa \end{array} \right)
& = & \left( \begin{array}{c}
1 \\ -\sqrt{\frac{E-m}{E+m}} \end{array} \right)
(kr)^{\gamma_\kappa-1} \frac{2^{\gamma_\kappa}e^{\pi \eta/2} \mid \Gamma(
\gamma_\kappa+i \eta) \mid}{\Gamma ( 2 \gamma_\kappa +1)} \nonumber \\
& & \left( \begin{array}{c} 
\mbox{Re} \\ \mbox{Im} \end{array} \right)
[(\gamma_\kappa+i \eta) e^{i\varphi} e^{-ikr} F(\gamma_\kappa + 1+ i \eta,
2 \gamma_\kappa+1; 2ikr)] ,  \label{radialfunctions}
\end{eqnarray}
where
\begin{displaymath}
\gamma_\kappa=\sqrt{\kappa^2 - \xi^2} \quad , \quad
\eta=\frac{\xi E}{k},
\end{displaymath}
\begin{equation}
e^{2i\varphi}=\frac{-\kappa +i\eta m/E}{\gamma_\kappa+i \eta} \quad , \quad
k=\sqrt{E^2-m^2}. \label{definitions}
\end{equation}
$\varphi$ is positive for $\mbox{arg} \, e^{2i\varphi} \in [0,\pi]$
and negative for  $\mbox{arg} \, e^{2i\varphi} \in (-\pi,0]$.
These Coulomb wave functions have the asymptotic forms
($r \rightarrow \infty$)
\begin{equation}
g_\kappa(r) \sim \frac{1}{kr} \cos(kr+\eta \log 2 k r - (l+1) \frac{\pi}{2}
+\delta_\kappa) \, , \label{asy1}
\end{equation}
\begin{equation}
f_\kappa(r) \sim -\sqrt{\frac{E-m}{E+m}} \frac{1}{kr}
\sin(kr+\eta \log 2 k r - (l+1) \frac{\pi}{2}
+\delta_\kappa) \, , \label{asymp} 
\end{equation}
where
\begin{displaymath}
\delta_\kappa=\frac{1}{2} \mbox{arg} \frac{-\kappa+\frac{i \eta m}{E}}
{\gamma_\kappa+i \eta}
-\mbox{arg} \, \Gamma (\gamma_\kappa+i \eta) - \gamma_\kappa
\frac{\pi}{2} + (l+1) \frac{\pi}{2}:=
\end{displaymath}
\begin{equation}
\varphi-\mbox{arg} \, \Gamma (\gamma+i \eta) + (l+1-\gamma) \frac{\pi}{2}.
\end{equation}
In the limiting case $Z \rightarrow 0$, we have
\begin{equation}
g_\kappa(r) \rightarrow - \mbox{sgn} (\kappa) j_l (kr) \, , \quad
f_\kappa(r) \rightarrow -\sqrt{\frac{E-m}{E+m}} j_{l'}
(kr), \label{freefield}
\end{equation}
where $j_l(kr)=\sqrt{\frac{\pi}{2kr}} J_{l+1/2}(kr)$ are
spherical Bessel functions of the first kind, and $l'=j-\frac{1}{2}
\mbox{sgn}(\kappa)=l(-\kappa)$.
Replacing $\gamma_\kappa$ by $-\gamma_\kappa$ in eq. (\ref{radialfunctions})
and correspondingly in eq. (\ref{definitions})
leads to the irregular solutions $g_\kappa^i$,
$f_\kappa^i$. Their asymptotic forms are also given by
eqns. (\ref{asy1},\ref{asymp}),
if one replaces the expression for the phase shift
$\delta_\kappa(\gamma_\kappa)$ of the regular solutions
by $\delta_\kappa' (\gamma_\kappa)=\delta_\kappa(-\gamma_\kappa)$.

The calculation of the wave functions $g_{\kappa}$ and $f_{\kappa}$
is most simply performed by using the real series expansion
\begin{equation}
\left( \begin{array}{c} g_{\kappa} \\ f_{\kappa} 
\end{array} \right) = (kr)^{\gamma_\kappa-1}
\sum_{n=0}^{\infty} \left( \begin{array}{c} a_{\kappa;n} \\
b_{\kappa;n} \end{array} \right) (kr)^n \, . \label{series}
\end{equation}
Inserting the series expansion (\ref{series})
into eq. (\ref{cdg}) leads to the coupled recursion relations
\cite{TBR83}
\begin{eqnarray}
(n+1)(2 \gamma_\kappa+n+1)k a_{\kappa;n+1}+\xi(E-m)
a_{\kappa;n} & & \nonumber \\ 
-(\gamma_\kappa+n+1-\kappa)(E+m)b_{\kappa;n} & = & 0 \, , \label{rec1}
\end{eqnarray}
\begin{eqnarray}
(n+1)(2 \gamma_\kappa+n+1)k b_{\kappa;n+1} + \xi(E+m)
b_{\kappa;n} & & \nonumber \\
+ (\gamma_\kappa+n+1+\kappa)(E-m) a_{\kappa;n} & = & 0 \, . \label{rec2}
\end{eqnarray}
The series expansion (\ref{series}) can be shown to converge
for all values of $r$.

We give here some details for the derivation of the
recursion relations. From
\begin{equation}
\frac{d}{dr} g(r)=(E+m+\xi/r) f(r) -\frac{\kappa+1}{r} g(r)
\end{equation}
one readily derives
\begin{displaymath}
(\gamma-1) (kr)^{\gamma-1} k \sum
\limits_{n=1}^{\infty} a_n (kr)^{n-1} + (\gamma-1) (kr)^{\gamma-1} 
\frac{a_0}{r} +  (kr)^{\gamma-1}  k
\sum \limits_{n=1}^{\infty} n a_n (kr)^{n-1} =
\end{displaymath}
\begin{displaymath}
(E+m) (kr)^{\gamma-1} \sum \limits_{n=0}^{\infty} b_n (kr)^{n} +
\xi (kr)^{\gamma-1} k \sum \limits_{n=1}^{\infty} b_n (kr)^{n-1}+
\xi (kr)^{\gamma-1} \frac{b_0}{r}
\end{displaymath}
\begin{equation}
-(\kappa+1) (kr)^{\gamma-1} k \sum \limits_{n=1}^{\infty} a_n (kr)^{n-1}
-(\kappa+1) (kr)^{\gamma-1} \frac{a_0}{r}, \label{insert1}
\end{equation}
where we have omitted the index $\kappa$ for notational
convenience. Comparing the lowest order terms $\sim r^{\gamma-2}$
immediately leads to the starting relation
\begin{equation}
b_0=\frac{\kappa+\gamma}{\xi} a_0=\frac{\xi}{\kappa-\gamma} a_0. \label{starting}
\end{equation}
From
\begin{equation}
\frac{d}{dr} f(r)=-(E-m+\xi/r) g(r) +\frac{\kappa-1}{r} f(r)
\end{equation}
one obtains
\begin{displaymath}
(\gamma-1) (kr)^{\gamma-1} k \sum
\limits_{n=1}^{\infty} b_n (kr)^{n-1} + (\gamma-1) (kr)^{\gamma-1} 
\frac{b_0}{r} +  (kr)^{\gamma-1} k
\sum \limits_{n=1}^{\infty} n b_n (kr)^{n-1} =
\end{displaymath}
\begin{displaymath}
-(E-m) (kr)^{\gamma-1} \sum \limits_{n=0}^{\infty} a_n (kr)^{n} -
\xi (kr)^{\gamma-1} k \sum \limits_{n=1}^{\infty} a_n (kr)^{n-1}-
\xi (kr)^{\gamma-1} \frac{a_0}{r}
\end{displaymath}
\begin{equation}
+(\kappa-1) (kr)^{\gamma-1} k \sum \limits_{n=1}^{\infty} b_n (kr)^{n-1}
+(\kappa-1) (kr)^{\gamma-1} \frac{b_0}{r}. \label{insert2}
\end{equation}
Comparing again the lowest order terms $\sim r^{\gamma-2}$
leads again to the starting relation (\ref{starting}).
Taking into account the higher order terms
$\sim (kr)^{\gamma-1} (kr)^{n-1}$ in eqns. (\ref{insert1},\ref{insert2})
leads to
\begin{equation}
(\gamma-1) k a_n +n k a_n -(E+m) b_{n-1} -\xi k b_n + (\kappa+1) k a_n=0 \,
\label{comp1}
\end{equation}
\begin{equation}
(\gamma-1) k b_n +n k b_n +(E-m) a_{n-1} +\xi k a_n - (\kappa-1) k b_n=0.
\label{comp2}
\end{equation}
Replacing $b_n$ from eq. (\ref{comp1})
\begin{equation}
b_n=\frac{\gamma+\kappa+m}{\xi} a_n - \frac{E+m}{\xi k} b_{n-1}
\end{equation}
in eq. (\ref{comp2}) gives
\begin{equation}
n(2 \gamma+n) k a_n + \xi (E-m) a_{n-1} - (\gamma+n-\kappa)
(E+m) b_{n-1}=0,
\end{equation}
which is equivalent to eq. (\ref{rec1}). Recursion relation
(\ref{rec2}) is obtained analogously.

An incident (outgoing) electron with asymptotic momentum $\vec{k}$,
energy $E$ and polarization $\tau$ is given by a linear combination
of the $\psi_\kappa^\mu$
\begin{equation}
\psi_\tau=4 \pi \sqrt{\frac{E+m}{2E}} \sum \limits_{\kappa \mu}
e^{\pm i \delta_\kappa} i^l \langle l \, (\mu-\tau) \, \frac{1}{2} \,
\tau  \, | \, j \, \mu \rangle (Y_l^{\mu-\tau} (\hat{\bf{k}}))^*
\psi_\kappa^\mu(\vec{r}) . \label{expansion}
\end{equation}
It is instructive to consider, e.g., the first component of the
Dirac spinor $\psi_\tau$
for an electron incident along the z-axis with
spin in the same direction. Then, the term $(Y_l^{\mu-\tau}
(\hat{\bf{k}})^*$
is only non-zero when $\mu=\tau=1/2$, and we have
\begin{equation}
Y_l^0 (\vartheta,\varphi)  = \sqrt{\frac{2l+1}{4 \pi}}
P_l(\cos \vartheta), \quad
Y_l^0 (\vartheta=0,\varphi=0)  =  \sqrt{\frac{2l+1}{4 \pi}}
\end{equation}

Therefore, the
first component of $\psi_\tau$ is given by
\begin{equation}
\psi_\tau^1=\sqrt{\frac{E+m}{2E}} \sum \limits_{\kappa}
(2l+1) e^{i \delta_\kappa} i^l \langle l \, 0 \, \frac{1}{2}
\frac{1}{2}  \, | \, j \, \frac{1}{2} \rangle^2
g_\kappa(r)  P_l(\cos \vartheta). \label{expa}
\end{equation}
A straightforward calculation shows that
\begin{equation}
\langle l \, 0 \, \frac{1}{2}
\frac{1}{2}  \, | \, j \, \frac{1}{2} \rangle^2=\frac{\kappa}{2 \kappa +1}.
\end{equation}
Furthermore, the asymptotic behavior of $g_\kappa$ is given by
\begin{equation}
g_\kappa(r) \sim \frac{1}{kr} \cos(kr+\eta \log 2 kr -\mbox{arg} \,
\Gamma(\gamma+i \eta)+\frac{1}{2} \mbox{arg} \,
\frac{-\kappa+i \eta m/E}{\gamma+i \eta} -\gamma \frac{\pi}{2}).
\end{equation}
We consider now the limit $Z \rightarrow 0$ for this asymptotic
expression. For $Z \rightarrow 0$ we have also $\eta \rightarrow 0$,
i.e. $\mbox{arg} \, \Gamma(\gamma+i \eta) \rightarrow 0$ and
$\gamma \rightarrow |\kappa|$.
From
\begin{equation}
e^{2 i \varphi} =\frac{-\kappa+i \eta m/E}{\gamma+i \eta}=
\frac{-\kappa \gamma+\eta m/E}{\gamma^2+\eta^2}+ i
\frac{\eta \kappa+\eta m /E}{\gamma^2+\eta^2}
\end{equation}
we see that the argument of $\varphi$ approaches
$\pi/2$ for $\eta \rightarrow 0$ and $\kappa>0$, whereas for
$\kappa <0$ we have $\varphi \rightarrow 0$.\\
This shows that the asymptotic behavior of the $g_\kappa$
for $Z \rightarrow 0$ is given by
\begin{equation}
g_\kappa (r) \sim \frac{1}{kr} \cos(kr-\frac{\pi l}{2}+\mbox{sgn}
(\kappa) \frac{\pi}{2})=-\mbox{sgn}(\kappa) \sin
(kr-\frac{\pi l}{2}),
\end{equation}
in accordance with eq. (\ref{freefield}) which states that the
free-field solutions of $g_\kappa$ are given by
$g_\kappa(r) = -\mbox{sgn}(\kappa) j_l (kr)$.
\noindent Therefore, the terms in the expansion (\ref{expa}) for
$\kappa > 0$ become ($l=\kappa$, $\delta_\kappa \rightarrow \pi$)
\begin{equation}
-\sqrt{\frac{E+m}{2E}} \sum \limits_{\kappa >0} l i^l e^{i \pi}
\frac{1}{kr} j_l(kr) P_l (\cos \vartheta)
\end{equation}
and for $\kappa<0$ ($l=-\kappa-1$, $\delta_\kappa \rightarrow 0$)
\begin{equation}
\sqrt{\frac{E+m}{2E}} \sum \limits_{\kappa < 0} (l+1) i^l
\frac{1}{kr} j_l(kr) P_l (\cos \vartheta),
\end{equation}
i.e. we obtain the partial wave expansion of a plane wave,
and the normalization is such that the full free spinor
for arbitrary momentum $\vec{k}$ and helicity $\zeta$ is given by
\begin{equation}
u_\tau = \sqrt{\frac{E+m}{2E}} \left(
\begin{array}{c}
\chi_{\tau} \\
\frac{\vec{\sigma} \vec{k}}{E+m} \chi_\tau
\end{array} \right) .
\end{equation}

For the case of a realistic nuclear electrostatic potential,
analytic expression for the radial functions are no longer
available. Therefore, we calculated the radial wave functions
by numerical integration according to the method described in
appendix 3 of \cite{Yennie54}. Outside the nuclear charge distribution
(i.e. for $r > 14$ fm in our actual calculations), the
electrostatic potential is a Coulomb potential, and therefore
the radial functions $G_\kappa$ obtained from the numerical integration
can be written as a linear combination of regular and
irregular solutions of the Dirac equation with a Coulomb
potential
\begin{equation}
G_\kappa=c_\kappa g_\kappa + d_\kappa g_\kappa^i.
\end{equation}
Since the asymptotic behavior of the regular and irregular
radial functions is given by
\begin{equation}
g_\kappa(r) =  \frac{1}{kr} \sin(kr+\eta \log 2 k r - l \frac{\pi}{2}
+\delta_\kappa) 
\end{equation}
\begin{displaymath}
g_\kappa^i(r)  =  \frac{1}{kr} \sin(kr+\eta \log 2 k r - l \frac{\pi}{2}
+\delta'_\kappa)
\end{displaymath}
\begin{equation}
=  \frac{1}{kr} \sin(kr+\eta \log 2 k r - l \frac{\pi}{2}
+ \delta_\kappa +  (\delta'_\kappa- \delta_\kappa)),
\end{equation}
and the asymptotic behavior of $G_\kappa$ is described by
the phase shift $\Delta_\kappa$ via
\begin{equation}
G_\kappa (r) \sim  \frac{\lambda}{kr} \sin(z+\Delta_\kappa)
\end{equation}
with $z=kr+\eta \log 2 k r - l \frac{\pi}{2}
+\delta_\kappa$, such that we obtain for $\Delta_\kappa$
\begin{displaymath}
\lambda \sin(z+\Delta_\kappa) = \lambda \sin(z) \cos(\Delta_\kappa)
+ \lambda \sin(\Delta_\kappa) \cos(z)
\end{displaymath}
\begin{displaymath}
=c_\kappa \sin(z) + d_\kappa \sin(z+\delta'_\kappa-\delta_\kappa)
\end{displaymath}
\begin{equation}
= c_\kappa \sin(z) +d_\kappa \sin(z) \cos(\delta'_\kappa-\delta_\kappa)
+d_\kappa \sin(\delta'_\kappa-\delta_\kappa) \cos(z) ,
\end{equation}
the relations
\begin{equation}
\lambda \cos{\Delta_\kappa}=
c_\kappa + d_\kappa \cos(\delta'_\kappa-\delta_\kappa) \quad ,
\quad \lambda \sin(\Delta_\kappa) = d_\kappa
\sin(\delta'_\kappa-\delta_\kappa)
\end{equation}
and therefore
\begin{equation}
\tan{\Delta_\kappa}=
\frac{\sin(\delta'_\kappa-\delta_\kappa)}{c_\kappa / d_\kappa
+\cos (\delta'_\kappa-\delta_\kappa)},
\end{equation}
which fix uniquely the phase shift $e^{i \Delta_\kappa}$.
The radial function $G_\kappa$ (and the
corresponding $F_\kappa$ for the lower
spinor components) obtained from the numerical
integration procedure must subsequently be multiplied
by a factor $\lambda^{-1}$, where
\begin{equation}
\lambda^2=(\lambda \cos{\Delta_\kappa})^2+(\lambda \sin{\Delta_\kappa})^2=
(c_\kappa^2 +d_\kappa^2+2 c_\kappa d_\kappa \cos
{\Delta_\kappa})
\end{equation}
such that
\begin{equation}
{\tilde{G}_\kappa}=\frac{1}{\lambda} G_\kappa (r) \sim 
\frac{1}{kr} \sin(kr+\Delta_\kappa)
\end{equation}
is correctly normalized.

The expansion for an incoming wave scattering off a spherically symmetric
nuclear charge distribution is finally given by
\begin{equation}
\psi_\tau=4 \pi \sqrt{\frac{E+m}{2E}} \sum \limits_{\kappa \mu}
e^{i \delta_\kappa} e^{i \Delta_\kappa}
i^l \langle l \, (\mu-\tau) \, \frac{1}{2} \,
\tau  \, | \, j \, \mu \rangle Y_l^{\mu-\tau*}  (\hat{\bf{k}})
\psi_\kappa^\mu (\vec{r}) , \label{expansion2}
\end{equation}
where $\psi_\kappa^\mu$ is defined according to eq. (\ref{sphericalspinor})
with the radial functions $g_\kappa$ and $f_\kappa$ replaced by
$\tilde{G}_\kappa$ and $\tilde{F}_\kappa$.

\end{document}